\documentclass[english,prd,twocolumn,superscriptaddress,floatfix,nofootinbib,preprintnumbers,eqsecnum]{revtex4}
\usepackage{graphicx}
\usepackage{amsmath,amsthm,amssymb}
\usepackage{bm}

\usepackage{amsfonts}
\usepackage{dcolumn}
\usepackage{hyperref}

\def\be{\begin{equation}}
\def\ee{\end{equation}}
\def\ba{\begin{eqnarray}}
\def\ea{\end{eqnarray}}
\def\bs{\begin{subequations}}
\def\es{\end{subequations}}

\usepackage{color}

\makeatother

\usepackage{babel}
\makeatother

\begin{document}

\preprint{YITP-15-72}

\title{Existence and disappearance of conical singularities 
in Gleyzes-Langlois-Piazza-Vernizzi theories}

\author{Antonio De Felice}

\affiliation{Yukawa Institute for Theoretical Physics,
Kyoto University, 606-8502, Kyoto, Japan}

\author{Ryotaro Kase}

\affiliation{Department of Physics, Faculty of Science,
Tokyo University of Science, 1-3,
Kagurazaka, Shinjuku, Tokyo 162-8601, Japan}

\author{Shinji Tsujikawa}

\affiliation{Department of Physics, Faculty of Science, 
Tokyo University of Science, 1-3,
Kagurazaka, Shinjuku, Tokyo 162-8601, Japan}

\begin{abstract}

  In a class of Gleyzes-Langlois-Piazza-Vernizzi (GLPV) theories, we
  derive both vacuum and interior Schwarzschild solutions under the
  condition that the derivatives of a scalar field $\phi$ with respect
  to the radius $r$ vanish.  If the parameter $\alpha_{\rm H}$
  characterizing the deviation from Horndeski theories approaches a
  non-zero constant at the center of a spherically symmetric body, we
  find that the conical singularity arises at $r=0$ with the Ricci
  scalar given by $R=-2\alpha_{\rm H}/r^2$.  This originates from
  violation of the geometrical structure of four-dimensional curvature
  quantities.  The conical singularity can disappear for the models in
  which the parameter $\alpha_{\rm H}$ vanishes in the limit that
  $r \to 0$. We propose explicit models without the conical
  singularity by properly designing the classical Lagrangian in such a
  way that the main contribution to $\alpha_{\rm H}$ comes from the
  field derivative $\phi'(r)$ around $r=0$.  We show that the
  extension of covariant Galileons with a diatonic coupling allows for
  the recovery of general relativistic behavior inside a so-called
  Vainshtein radius. In this case, both the propagation of a fifth
  force and the deviation from Horndeski theories are suppressed
  outside a compact body in such a way that the model is compatible
  with local gravity experiments inside the solar system.

\end{abstract}

\date{\today}


\maketitle

\section{Introduction}
\label{intro} 

Over the past few decades, there have been numerous attempts for
extending General Relativity (GR) to more general gravitational
theories \cite{review}.  This is motivated by the ultra-violet
completion of gravity \cite{string,Horava} or by the observational
evidence of early and late phases of cosmic acceleration
\cite{SNIa,Planck}.  In particular, the problem of dark energy implies
that the today's acceleration of the Universe may be related to some
infra-red modification of gravity.

If we extend GR in such a way that only one additional scalar degree
of freedom (DOF) arises, Horndeski theories \cite{Horndeski} are known
as the most general scalar-tensor theories with {\it second-order}
equations of motion \cite{Horndeski2}.  Theories with derivatives
higher than second order can be prone to instabilities due to the
appearance of extra radiative DOF \cite{Ost}.  Nevertheless, it is
possible to generalize Horndeski theories to those with only one
propagating scalar DOF.  In Ho\v{r}ava-Lifshitz gravity \cite{Horava},
for example, addition of higher-order {\it spatial} derivatives does
not increase the number of the scalar DOF.

Another possibility for the extension of second-order gravitational
theories is to violate two conditions Horndeski theories obey
\cite{GLPV}.  One of such conditions is related to a geometric
modification of the Einstein-Hilbert Lagrangian
$L_{\rm EH}=M_{\rm pl}^2 R/2$, where $R$ is the four-dimensional Ricci
scalar and $M_{\rm pl}$ is the reduced Planck mass.  In terms of the
3+1 Arnowitt-Deser-Misner (ADM) decomposition of space-time
\cite{ADM}, this Lagrangian can be expressed in the form
$L_{\rm EH}=A_4(K^2-K_{\mu\nu}K^{\mu \nu})+B_4 {\cal R}$ with
$-A_4=B_4=M_{\rm pl}^2/2$, where $K$ and ${\cal R}$ are the traces of
three-dimensional extrinsic and intrinsic curvatures respectively.  On
the isotropic cosmological background Horndeski theories contain the
same form of Lagrangian as $L_{\rm EH}$, but $A_4$ and $B_4$ are
functions of the scalar field $\phi$ and its kinetic term $X$.  These
functions satisfy the particular relation $A_4=2XB_{4,X}-B_4$
\cite{building}, where $B_{4,X} \equiv \partial B_4/\partial X$.

GLPV theories do not obey this particular relation as well as another condition associated with the Horndeski Lagrangian 
$L_5$ \cite{GLPV}. 
In the unitary gauge, where the constant field hypersurfaces are 
identified with the constant time hypersurfaces, the Hamiltonian analysis shows that GLPV theories possess only one scalar DOF \cite{Hami2}. 
The simplest class of GLPV theories corresponds to
functions $A_4$ and $B_4$ depending on $\phi$ alone, in which case the
difference between $-A_4$ and $B_4$ characterizes the deviation from
Horndeski theories.  This deviation gives rise to several interesting
observational signatures such as the mixing of scalar and matter
propagation speeds \cite{Gergely,GLPV,Kase14}, the large gravitational
slip parameter \cite{Koyama}, and the realization 
of weak gravity \cite{Tsuji15}.

For the consistency with local gravity experiments, the interaction with matter mediated by the scalar DOF should be suppressed inside the solar system \cite{Will}.  
In Horndeski theories, nonlinear scalar-field self interactions can screen the fifth force \cite{Kimura,KaseHo} 
through the Vainshtein mechanism \cite{Vainshtein}.  
In GLPV theories it was recently claimed that the
Vainshtein mechanism tends to break down around the vicinity of a compact object \cite{Koba,Sakstein,Mizuno}.  Whether this conclusion holds or not for all classes of GLPV theories remains to be an open question.

In this paper, we revisit the analysis of spherically symmetric
solutions in GLPV theories to study how much deviation from Horndeski
theories can be allowed. On using the background equations of motion
derived in Sec.~\ref{EOM}, we first obtain vacuum solutions around a
point source ($r=0$) in Sec.~\ref{vacuumsec} and show that the Ricci
scalar is given by $R=-2\alpha_{\rm H}/r^2$ in the absence of the
cosmological constant $\Lambda$.  Since $\alpha_{\rm H} \neq 0$ in
GLPV theories, there is a conical singularity at the origin.  In
Sec.~\ref{internalsec} we also derive solutions inside a compact
object for constant $\alpha_{\rm H}$ under the condition that radial
derivatives of $\phi(r)$ vanish everywhere and show that $R$ is again
divergent at $r=0$.  Since the boundary condition $\phi'(r)=0$ at
$r=0$ is most natural for the regularity of solutions, the conical
singularity inevitably arises for the theories in which
$\alpha_{\rm H}$ approaches a non-zero constant as $r \to 0$.

A possible way out for avoiding the conical singularity is to consider
the theories with a vanishing $\alpha_{\rm H}$ at the origin by
designing the Lagrangian such that the main contribution to
$\alpha_{\rm H}$ corresponds to $X$-dependent terms. 
The exact form of choice for the Lagrangian is essential to remove 
such a conical singularity, and the absence of a symmetry which 
preserves its form might be a problem, once one considers, 
e.g.,\ quantum corrections. In Sec.~\ref{conditionsec} we
discuss conditions for eliminating the conical singularity in the
framework of GLPV theories and present an explicit Lagrangian which
allows for this possibility.  In Sec.~\ref{modelsec} we derive the
field profile and gravitational potentials inside and outside the body
under the approximation of weak gravity for a concrete model without
the conical singularity.  We show that the Vainshtein mechanism is
efficient enough to suppress the fifth force and the deviation from
Horndeski theories.  We conclude in Sec.~\ref{consec}.

\section{Background equations of motion}
\label{EOM} 

We consider the theories described by the action \cite{GLPV}
\be
S=\int d^4 x \sqrt{-g} \sum^{4}_{i=2}L_i
+\int d^4 x \sqrt{-g}\,L_m (g_{\mu \nu},\Psi_m)\,,
\label{act}
\ee
where $g$ is the determinant of metric $g_{\mu \nu}$, and $L_m$ is the
Lagrangian of matter fields $\Psi_m$ with the energy-momentum tensor
$T^{\mu}_{\nu}={\rm diag}\,(-\rho_{m},P_{m},P_{m},P_{m})$.  We assume
that matter is minimally coupled to the metric $g_{\mu \nu}$.  The Lagrangians $L_{2},L_{3}$, and $L_{4}$ are given by
\ba
L_{2} &=&A_{2}(\phi ,X)\,,  
\label{L2} \\
L_{3} &=&\left( C_{3}+2XC_{3,X} \right)
\square \phi +XC_{3,\phi}\,,  
\label{L3} \\
L_{4} &=&B_{4} R-\frac{B_{4}+A_{4}}
{X}\left[ (\square \phi )^{2}-\nabla ^{\mu} 
\nabla ^{\nu}\phi \nabla_{\mu}\nabla_{\nu}\phi \right]  \nonumber \\
&&+\frac{2\left( B_{4}+A_{4}-2XB_{4,X} \right)}
{X^{2}} ( \nabla ^{\mu}\phi \nabla ^{\nu}\phi 
\nabla _{\mu}\nabla _{\nu}\phi\,
\square \phi \nonumber \\
& &-\nabla ^{\mu}\phi \nabla _{\mu}\nabla _{\nu}\phi 
\nabla _{\sigma}\phi
\nabla ^{\nu}\nabla ^{\sigma}\phi )\,, 
\label{L4} 
\ea
where $\nabla_{\mu}$ denotes covariant derivatives,
$A_2, C_3, A_4, B_4$ are functions of the scalar field $\phi$ and its kinetic energy 
$X=g^{\mu \nu}\nabla_{\mu}\phi \nabla_{\nu}\phi$.  
Introducing the function $F_4$ defined by 
\be
-F_4 X^2=B_4+A_4-2XB_{4,X}\,,
\ee
one can write the Lagrangian $L_4$ of the form \cite{Koba}
\ba
\hspace{-0.5cm}
L_{4} &=&B_{4} R-2B_{4,X}
\left[ (\square \phi )^{2}-\nabla ^{\mu} \nabla ^{\nu}\phi \nabla_{\mu}\nabla_{\nu}\phi \right]  \notag \\
\hspace{-0.5cm}
& &+F_{4}\epsilon^{\mu\nu\rho\sigma}\epsilon_{\mu'\nu'\rho'\sigma} 
\nabla^{\mu'}\phi\nabla_{\mu}\phi
\nabla^{\nu'}\nabla_{\nu}\phi
\nabla^{\rho'}\nabla_{\rho}\phi\,,
\ea
where $\epsilon_{\mu\nu\rho\sigma}$ 
is the totally antisymmetric Levi-Civita tensor. 
Horndeski theories satisfy the condition $F_4=0$, 
i.e., $B_4+A_4-2XB_{4,X}=0$, under
which the terms after the second lines 
of Eq.~(\ref{L4}) vanish. 

The full action of GLPV theories contains the Lagrangian $L_5$ associated with the Einstein tensor, 
but we do not take into account such a
contribution in this paper.  
In fact, inclusion of $L_5$ tends to prevent the
recovery of GR in the solar system even 
in Horndeski theories \cite{Kimura,KaseHo}. 

Using the 3+1 ADM formalism, we can construct several scalar
quantities such as $K=g^{\mu \nu}K_{\mu \nu}$ and
${\cal R}=g^{\mu \nu}{\cal R}_{\mu \nu}$ from 
the extrinsic curvature $K_{\mu \nu}$ and the intrinsic curvature ${\cal R}_{\mu \nu}$ \cite{building}. 
In the unitary gauge ($\phi=\phi(t)$), the Lagrangian 
$L=L_2+L_3+L_4$ is equivalent to 
$L=A_2+A_3K+A_4(K^2-K_{\mu\nu}K^{\mu \nu})+B_4 {\cal R}$ 
with the relation \cite{GLPV,Gergely2,KTD15}
\be
A_{3}=2|X|^{3/2} \left( C_{3,X}+\frac{B_{4,\phi}}{X} 
\right)\,,
\label{A3def}
\ee
where the sign of $X$ is different depending on the given space-time. 
On the spherically symmetric 
and static background we have that $X>0$, whereas 
on the isotropic and homogenous background $X<0$. 

The deviation from Horndeski theories can be quantified by the parameter \cite{GLPV}
\be
\alpha_{\rm H} \equiv \frac{2XB_{4,X}-B_4}
{A_4}-1=\frac{F_4X^2}{A_4}\,.
\ee
We also define \cite{Bellini}
\be
\alpha_{\rm t} \equiv -\frac{B_4}{A_4}-1\,.
\ee
For linear perturbations on the spherically symmetric background,
$\alpha_{\rm t}$ is related to the tensor propagation speed squared
$c_{\rm t}^2$, as $\alpha_{\rm t}=1/c_{\rm t}^2-1$
\cite{Moto,Gergely2}.  Then, $\alpha_{\rm t}$ characterizes the
deviation of $c_{\rm t}^2$ from 1.  If the function $B_4$ depends on
$\phi$ alone, $\alpha_{\rm H}$ is equivalent to $\alpha_{\rm t}$.

We consider the spherically symmetric background described by the
metric
\be
ds^{2}=-e^{2\Psi(r)}dt^{2}+e^{2\Phi(r)}dr^{2}
+r^{2} (d\theta^{2}+\sin^{2}\theta\, 
d\varphi^{2})\,,
\label{line}
\ee
where the gravitational potentials $\Psi(r)$ and $\Phi(r)$ are
functions of the distance $r$ from the center of symmetry.  Variation
of the action (\ref{act}) leads to the equations of motion
\ba
&&
\left( \frac{4e^{-2\Phi}A_4}{r}-{\cal C}_1+4{\cal C}_2 
\right)\Phi'-A_2+{\cal C}_3-{\cal C}_4 \nonumber \\
&&
-\frac{2A_4}{r^2} \left( e^{-2\Phi}-1-\alpha_{\rm t} \right)
=-\rho_m\,,\label{eq:00}\\
&&\left( \frac{4e^{-2\Phi}A_4}{r}-{\cal C}_1+4{\cal C}_2 
\right)\Psi'+A_2-2XA_{2,X}-\frac{2{\cal C}_1}{r} \nonumber \\
&&
+\frac{2}{r^2} \left[  A_4 (e^{-2\Phi}-1-\alpha_{\rm H})
+r{\cal C}_2 \right]
=-P_m\,,\label{eq:11}\\
&& 2e^{-2\Phi}A_4 \left( \Psi''+{\Psi'}^2 \right)+
\left( \frac{2e^{-\Phi}A_4}{r}+\frac12 {\cal C}_4 r \right) 
\Psi'  \nonumber \\
&&
+\left[ {\cal C}_1-2 \left( \frac{e^{-2\Phi}A_4}{r}
+{\cal C}_2 \right) \left(1+r\Psi' \right) \right]\Phi' 
\nonumber \\
& &
+A_2-{\cal C}_3+\frac12 {\cal C}_4
=-P_m\,,\label{eq:22} \\
& &
P_m'+\Psi' \left( \rho_m+P_m \right)=0\,,
\label{eq:33}
\ea
where $X=e^{-2\Phi} \phi'^2$, a prime represents a derivative with
respect to $r$, and
\ba
& &
{\cal C}_1 \equiv 2e^{-\Phi}XA_{3,X}\,,\qquad
{\cal C}_2 \equiv \frac{2e^{-2\Phi}XA_{4,X}}{r}\,,
\nonumber \\
& &
{\cal C}_3 \equiv e^{-\Phi}\phi' \left( A_{3,\phi}+
2e^{-2\Phi}\phi'' A_{3,X} \right)\,,\nonumber \\
& &
{\cal C}_4 \equiv \frac{4e^{-2\Phi}\phi' \left( A_{4,\phi}+
2e^{-2\Phi}\phi'' A_{4,X} \right)}{r}\,.
\label{Cidef}
\ea
In GR we have $-A_4 =B_4 = M_{\rm pl}^2/2$ and
$\alpha_{\rm H}=\alpha_{\rm t}=0$.  In GLPV theories, the parameter
$\alpha_{\rm H}$ gives rise to a contribution comparable to the
dominant term $A_4\Phi/r^2$ in Eq.~(\ref{eq:11}) for
$|\alpha_{\rm H}|$ larger than the order of $|\Phi|$.

In the following we will assume that both $A_4$ and $B_4$ 
also remain finite as $r \to 0$. For this class of theories 
(without shift symmetry), on a static spherically symmetric background, 
the field $\phi$ is required to be only $r$-dependent, and 
$\phi'(r \to 0)=0$ for the regularity in compact objects. 
In this case, we have that $X \to 0$ at the origin.

It should be noticed that the action (\ref{act}) remains 
finite in the limit $X \to 0$. In fact, for analytic profiles 
for the field, one can verify that
$\sqrt{-g} L_4\to\mathrm{constant}$ around the origin.
This should not come as a surprise as, in fact, the same 
limit is well-defined also for Horndeski theories, 
whose Lagrangian reduces to the first line of Eq.~(\ref{L4}). 
We also note that the equations of motion 
(\ref{eq:00})-(\ref{eq:22}) do not contain the $X$-dependent 
term in the denominators. 
It is clear that finite values of $A_4$ and $B_4$
are allowed as long as $\Phi',\Psi'\to0$, 
as requested for standard boundary conditions 
for compact objects.

\section{Vacuum Schwarzschild solutions}
\label{vacuumsec} 

First of all, we derive exact solutions to Eqs.~(\ref{eq:00}) and
(\ref{eq:11}) in the absence of matter for a point source with mass
${\cal M}$. Since the field derivatives $\phi'(r)$ and $\phi''(r)$
vanish in this case, the terms ${\cal C}_{i}$ defined in
Eq.~(\ref{Cidef}) do not contribute to the equations of motion.  The
terms $A_4$ and $B_4$ are regarded as constants, with the particular
relation $\alpha_{\rm H}=\alpha_{\rm t}=-B_4/A_4-1$.  The term $A_2$
is related to the cosmological constant $\Lambda~(>0)$, as
$A_2=-\Lambda$.  Then, Eqs.~(\ref{eq:00}) and (\ref{eq:11}) reduce to
\ba
& &
\frac{4e^{-2\Phi}A_4}{r}\Phi'+\Lambda
-\frac{2A_4}{r^2} \left( e^{-2\Phi}-1-\alpha_{\rm H} \right)=0\,,\\
& &
\frac{4e^{-2\Phi}A_4}{r}\Psi'-\Lambda
+\frac{2A_4}{r^2}(e^{-2\Phi}-1-\alpha_{\rm H})=0\,. 
\ea
The solutions to these equations are given by 
\ba
e^{2\Phi} &=& 
\left( 1+\alpha_{\rm H}+\frac{\Lambda}{6A_4}r^2
+\frac{c_1}{2A_4r} \right)^{-1}\,,
\label{Phiso}\\
e^{2\Psi} &=&
-6A_4c_2 \left( 1+\alpha_{\rm H}+\frac{\Lambda}{6A_4}r^2
+\frac{c_1}{2A_4r} \right)\,,
\label{Psiso}
\ea
where $c_1$ and $c_2$ are integration constants.  
The deviation from Horndeski theories leads to the contribution
to gravitational potentials.  
The cosmological constant gives rise to the contribution
$\Lambda r^2/(6A_4)$, as it happens for the Schwarzschild-de Sitter
solution. The two coefficients $c_1$ and $c_2$ also arise in GR.
The constant $c_1$ is chosen as $c_1=-4A_4G{\cal M}$ to obtain
the standard term $-2G{\cal M}/r$ in 
Eqs.~(\ref{Phiso}) and (\ref{Psiso}).  
To recover the usual relation $e^{2\Psi}=e^{-2\Phi}$, 
the constant $c_2$ is chosen to be $c_2=-1/(6A_4)$.  
In this way, we have fixed two freedoms (Schwarzschild 
mass ${\cal M}$ and the time reparametrization) as 
in the context of GR.

Then, we obtain the following solution
\be
e^{2\Psi}=e^{-2\Phi}=1+\alpha_{\rm H}
+\frac{\Lambda}{6A_4}r^2-\frac{2G{\cal M}}{r}\,.
\label{vasch}
\ee

This is the extension of the Schwarzschild-de Sitter solution with the
additional factor $\alpha_{\rm H}$.  Since
$g^{tt}R_{tt}=g^{rr}R_{rr}=-\Lambda/(2A_4)$ and
$g^{\theta \theta}R_{\theta \theta}= g^{\varphi \varphi}R_{\varphi
  \varphi}=-\Lambda/(2A_4) -\alpha_{\rm H}/r^2$,
the Ricci scalar is given by
\be
R=-\frac{2\Lambda}{A_4}-\frac{2\alpha_{\rm H}}{r^2}\,,
\label{Rva}
\ee
which diverges at $r=0$ for $\alpha_{\rm H} \neq 0$.  
Provided that $\alpha_{\rm H} \neq 0$ for $r \to 0$,
the divergence of $R$ occurs independent of the choices 
of $c_1$ and $c_2$, so it cannot be eliminated by the change
of boundary conditions.

The divergence of curvature at $r=0$ can be interpreted as a conical
singularity originating from the $\theta, \varphi$ contributions to
$R$.  This singularity cannot be eliminated even for $\Lambda \to 0$
and ${\cal M} \to 0$.  In this limit, the three-dimensional spatial
metric is given by
$ds_{(3)}^2=(1+\alpha_{\rm H})^{-1}dr^2+ r^2 (d\theta^2+\sin^2
\theta\,d\varphi^2)$.
Defining $\hat{r}=r/\sqrt{1+\alpha_{\rm H}}$, the two-dimensional
metric in the $\theta=\pi/2$ plane is of the form
$ds_{(2)}^2=d\hat{r}^2+\hat{r}^2 d\hat{\varphi}^2$, where
$\hat{\varphi}=\sqrt{1+\alpha_{\rm H}}\,\varphi$.  Since the angle
$\hat{\varphi}$ is not restricted to be between $0$ and $2\pi$, the
resulting space-time is a cone with a geodesically incomplete point at
$r=0$. 

To avoid such a conical singularity in the vacuum, we need to impose 
the condition $\alpha_{\rm H}=0$. In the next section we will show
that the same conclusion also holds in the presence of a matter source
for constant $\alpha_{\rm H}$.

\section{Interior Schwarzschild solutions for 
constant $\alpha_{\rm H}$}
\label{internalsec} 

We derive interior and exterior Schwarzschild solutions under the
conditions that the density $\rho_m$ inside a compact body is constant
and that $\phi'(r)=0$ and $\phi''(r)=0$ everywhere.  Again, we deal
with $\alpha_{\rm H}$ as a constant satisfying the relation
$\alpha_{\rm H}=\alpha_{\rm t}=-B_4/A_4-1$.

Integration of Eq.~(\ref{eq:33}) gives 
\be
P_m=-\rho_m+\mathcal{B} e^{-\Psi}\,,
\label{Pmso}
\ee
where $\mathcal{B}$ is a constant.  Solving Eq.~(\ref{eq:00}) for
$\Phi$, we obtain the same solution as Eq.~(\ref{Phiso}) with the
replacement $\Lambda \to \Lambda+\rho_m$.  For the regularity of
metric at $r=0$ we need to choose $c_1=0$, so the solution reduces to
\be
e^{2\Phi}= \left( 1+\alpha_{\rm H}-{\cal A} r^{2} \right)^{-1}\,,
\label{Phiin1}
\ee
where
\be
{\cal A} \equiv -\frac{\Lambda+\rho_m}{6A_4}\,.
\ee
There exists the term $\alpha_{\rm H}$ in Eq.~(\ref{Phiin1}),
which cannot be eliminated by any boundary condition.  
As we will see below, this again leads to the conical singularity at $r=0$.

Solving Eq.~(\ref{eq:11}) for $\Psi$, we obtain
\begin{equation}
e^{\Psi}=\frac{3\mathcal{B}}{2(\Lambda+\rho_m)}
-\mathcal{D}\sqrt{1+\alpha_{\rm H}-\mathcal{A}r^{2}}\,,
\label{Psiin1}
\end{equation}
where $\mathcal{D}$ is a constant.  There are two free integration
constants $\mathcal{B}$ and $\mathcal{D}$ in addition to the mass
${\cal M}$ of the compact object.  These constants can be fixed by
matching interior and exterior solutions of $\Phi$ and $\Psi$ at the
radius $r_0$ of the body determined by the condition $P_m(r_0)=0$.
Analogous to the derivation of Eq.~(\ref{vasch}), the exterior vacuum
solution for $r>r_0$ is given by
\be
e^{2\Psi}=e^{-2\Phi}=1+\alpha_{\rm H}
-\frac{\Lambda}{\Lambda+\rho_m}{\cal A}r^2
-\frac{2G{\cal M}}{r}\,.
\label{vasch2}
\ee
Matching the solutions at $r=r_0$, it follows that
\ba
{\cal B} &=& \rho_m \sqrt{1+\alpha_{\rm H}-{\cal A} r_{0}^{2}}\,,
\label{Bdef} \\
{\cal A} &=& \frac{2G{\cal M}}{r_0^3}
\frac{\Lambda+\rho_m}{\rho_m}\,,
\label{Mdef} \\
{\cal D} &=& \frac{\rho_m-2\Lambda}{2(\Lambda+\rho_m)}\,.
\label{Ddef}
\ea

After substitutions of Eqs.~(\ref{Bdef})-(\ref{Ddef}) into
Eqs.~(\ref{Phiin1}) and (\ref{Psiin1}), we obtain the interior
Schwarzschild solution with appropriate matching conditions.  Then, we
can find the behavior of three curvature scalars $R$,
$S \equiv R_{\mu \nu} R^{\mu \nu}$, and
$T \equiv R_{\mu \nu \lambda \rho} R^{\mu \nu \lambda \rho}$.  In the
limit where $r \to 0$, it follows that
\be
R \to -\frac{2\alpha_{\rm H}}{r^{2}}\,,\quad
S \to \frac{2\alpha_{\rm H}^{2}}{r^{4}}\,,\quad
T \to \frac{4\alpha_{\rm H}^{2}}{r^{4}}\,.
\ee
We have dropped next-to-leading order corrections, which appear as a
constant for $R$ and as the terms involving $\alpha_{\rm H}/r^2$ for
$S$ and $T$.  Taking into account these corrections, the Gauss-Bonnet
term $R_{\rm GB}^2=R^2-4S+T$ is proportional to 
$\alpha_{\rm H}/r^2$ around $r=0$. 
The integration constants have been fixed as
Eqs.~(\ref{Bdef})-(\ref{Ddef}), but the singular behavior of $R, S, T$
is independent of the choice of boundary conditions. Thus, for
$\alpha_{\rm H}={\rm constant}$, the conical singularity similar to
that discussed in Sec.~\ref{vacuumsec} is present at the center of 
a compact object.

It should be noted that the metric and matter fields are all regular 
at the origin, whereas the singularity appears only at the level 
of the Ricci scalar, i.e.,\ for tidal forces and geodesics deviation.
Although we have found the presence of this singularity for exact solutions (which always give a clear insight of theories), 
one may worry that it is merely a consequence of the fact 
we have assumed a specific constant field profile 
$\phi(r)=\mathrm{constant}$ everywhere.  
But this ``special'' solution is actually ``the''
solution which is approached at the origin for compact objects
(assuming the analytic profile for each field).

In fact, if the fields are analytic around $r=0$, then we can 
Taylor-expand them as: 
$\phi=\phi_c+\zeta_\phi\,r^2/2+\mathcal{O}(r^3)$, 
$\Phi=\Phi_c+\zeta_\Phi\,r^2/2+\mathcal{O}(r^3)$, 
$\Psi=\Psi_c+\zeta_\Psi\,r^2/2+\mathcal{O}(r^3)$,
and $\rho=\rho_c+\zeta_\rho\,r^2/2+\mathcal{O}(r^3)$. 
The coefficients $\zeta$'s are numerical constants 
which can be determined by the equations of motion 
or boundary conditions. 
Here, the crucial point to be shown is whether the variation 
of $\phi(r)$ induced by the term $\zeta_\phi\,r^2/2$ can 
modify the existence of the conical singularity.

We do not specify the functional forms of $A_2,A_3,A_4$, 
but the leading-order contributions to the terms $A_2$, 
$A_{3,X}$, $A_{3,\phi}$, $A_{4,X}$ and $A_{4,\phi}$ 
are regarded as constants.  
Using the above Taylor-expansions of $\phi$ and $\Phi$,  
the leading-order terms in Eq.~(\ref{Cidef}) are given, 
respectively, by ${\cal C}_1=\tilde{c}_1r^2$, 
${\cal C}_2=\tilde{c}_2r$, ${\cal C}_3=\tilde{c}_3 r$, 
and ${\cal C}_4=\tilde{c}_4$, where $\tilde{c}_i$'s 
are constants associated with $\zeta_\phi$. 
In the limit that $r \to 0$ the terms 
${\cal C}_1$, ${\cal C}_2$, and ${\cal C}_3$ vanish, 
so only the non-vanishing contributions to 
Eqs.~(\ref{eq:00})-(\ref{eq:22}) are $-{\cal C}_4$ 
in Eq.~(\ref{eq:00}) and ${\cal C}_4/2$ in 
Eq.~(\ref{eq:22}). These ${\cal C}_4$ terms
behave in the same way as another constant $A_2=-\Lambda$  
already studied above. This discussion shows that 
the variation of $\phi(r)$ around $r=0$ does not give 
rise to any new contribution to modify the existence 
of the conical singularity.

More concretely, one can verify from the integration of 
Eq.~(\ref{eq:00}) that the gravitational potential $e^{-2\Phi}$ 
acquires the terms proportional to $\tilde{c}_1r^5$, 
$\tilde{c}_2r^4$, $\tilde{c}_3r^3$, $\tilde{c}_4r^2$, 
and $\zeta_{\rho}r^4$ in addition to those 
given in Eq.~(\ref{vasch2}). 
In the limit that $r \to 0$ all these additional terms vanish, 
so the gravitational potential again approaches the value 
$e^{-2\Phi} \to 1+\alpha_{\rm H}$. 
Similarly the additional terms ${\cal C}_1$ and ${\cal C}_2$ 
appearing in Eq.~(\ref{eq:11}) only give rise to vanishing 
contributions to $\Psi$ as $r \to 0$, 
such that $e^{2\Psi} \to 1+\alpha_{\rm H}$ 
for $r \to 0$. In addition to the fact that the integrated 
solution (\ref{Pmso}) of Eq.~(\ref{eq:33}) is used 
in this procedure, the above solutions are consistent with Eq.~(\ref{eq:22}). Hence they are the solutions of the full 
equations of motion (\ref{eq:00})-(\ref{eq:33}) 
around $r=0$.
The existence of the term $\alpha_{\rm H}$ in 
$\Phi$ and $\Psi$ shows that 
the variation of the field $\phi(r)$ around $r=0$ cannot 
eliminate the conical singularity. 

The robustness of the behavior $e^{-2\Phi}=e^{2\Psi}=1+\alpha_{\rm H}$ 
for $r \to 0$ can be also confirmed by directly substituting the above 
expansions into the background equations of motion.
On expanding Eq.~(\ref{eq:00}) around the origin, 
we find that the first non-trivial contribution which needs to 
be canceled is given by 
\be
\frac2{r^2}\,[B_4(\phi_c,0)+e^{-2\Phi_c}\,A_4(\phi_c,0)]\,,
\ee
from which we obtain 
$e^{-2\Phi_c}=1+\alpha_{\rm H}(\phi_c,0)$. 
Applying the same procedure to Eq.~(\ref{eq:11}) with Eq.~(\ref{Pmso}), 
it follows that $e^{2\Psi_c}=1+\alpha_{\rm H}(\phi_c,0)$. 
Thus, the theories approaching a non-zero value of $\alpha_{\rm H}$ 
as $r \to 0$ lead to the appearance of the conical singularity.

It should be noticed that we have focused on static solutions and that 
we have not imposed the shift symmetry of the theory.
Abandoning any of the two assumptions could lead to a different 
behavior for the theory, and possibly, to the disappearance of the singularity. 
We leave this interesting topic for a future study.

\section{Conditions for existence and disappearance
of the conical singularity}
\label{conditionsec} 

The results in Sec.~\ref{vacuumsec} and \ref{internalsec} have been
derived under the conditions that (i) $\phi'(r)=0$ for $r \geq 0$ and
(ii) $\alpha_{\rm H}={\rm constant}$.  Since the boundary condition
$\phi'(0)=0$ is generic for the regularity of solutions
\cite{Khoury,DKT}, this is consistent with the assumption (i) around
$r=0$.  The remaining possibility for eliminating the conical
singularity is to construct models in which the parameter
$\alpha_{\rm H}$ approaches 0 as $r \to 0$.

Let us consider this second possibility in detail.  In general, around
the center of a compact object with spherical symmetry and staticity,
we should expect that $X \to 0$ for regularity.  Therefore, the
parameter $\alpha_{\rm H}$ may be expressed in the form
\begin{equation}
\alpha_{\rm H}=\alpha_{\rm H}(\phi_c,0)
+\mathcal{O}(X)\,.
\end{equation}
We need to impose $\alpha_{\rm H}(\phi_c,0)=0$ for removing the
conical singularity.  To fulfill this condition, there are two
possibilities: 1) to solve $\alpha_{\rm H}(\bar{\phi}_c,0)=0$ for
$\bar{\phi}_c$ if a solution (or even more than one) exists; 2) to
impose that the first term in the Taylor expansion (i.e.,\ the
constant term) in $X$ identically vanishes.

The first possibility cannot be viable in general, as this corresponds
to setting two different boundary conditions for the field at the
origin, that is $\phi'(r=0)=0$ (standard boundary condition),
$\phi(r=0)=\bar{\phi}_c$ (non-standard additional boundary
condition). Reflecting the fact that the field equation of motion is
of second order, these two boundary conditions completely fix the
dynamics of the field even at infinity. This would make the system
over-constrained, in general, and would reduce the freedom of
solutions.

The second possibility, which will be explored in the following,
consists of imposing that, in a Taylor expansion of $\alpha_{\rm H}$
around the origin, the constant term $\alpha_{\rm H}(\phi_c,0)$
identically vanishes for any $\phi_c$. In general this condition
cannot be imposed by any symmetry, so that we need to tune the choice
of the function $\alpha_{\rm H}$.  Moreover, it is not clear whether
such a functional form of $\alpha_{\rm H}$ is valid in the presence of
quantum corrections and whether such corrections spoil (or not) the
classical picture by making the conical singularity reappear.  The
analysis of this important issue is beyond the scope of our paper, but
we would like to discuss it in a future project.

To explore the second possibility further, we consider a canonical
scalar field $\phi$ without the potential, i.e., $A_2=-X/2$ and
$C_3=0$.  To go beyond the Horndeski domain, we need to choose
functions $A_4$ and $B_4$ which violate the condition
$A_4=2XB_{4,X}-B_4$.  One way is to extend the scalar-curvature
coupling appearing in Brans-Dicke theories \cite{Brans} or dilation
gravity \cite{Gas} to the forms $A_4=-M_{\rm pl}^2F_1(\phi)/2$ and
$B_4=M_{\rm pl}^2F_2(\phi)/2$, where $F_1(\phi)$ and $F_2(\phi)$ are
functions with respect to $\phi$.  Another way is to take into account
the $X$ dependence in the functions $A_4$ and $B_4$, as it happens for
covariant Galileons \cite{cova} and extended Galileons \cite{exga}.

To accommodate these two cases, we focus on the models characterized
by
\ba
& &A_2=-\frac12 X\,,\qquad C_3=0\,,\nonumber \\
& &A_4=-\frac12 M_{\rm pl}^2 F_1(\phi)+f_1(X)\,,\nonumber \\
& &B_4=\frac12 M_{\rm pl}^2 F_2(\phi)+f_2(X)\,,
\label{modelfun}
\ea
where $f_1(X)$ and $f_2(X)$ are functions of $X$.  From
Eq.~(\ref{A3def}) the function $A_3$ is given by
\be
A_3=M_{\rm pl}^2 \sqrt{X}F_{2,\phi}(\phi)\,.
\label{modelA3}
\ee
The parameters $\alpha_{\rm H}$ and 
$\alpha_{\rm t}$ read
\ba
\hspace{-0.5cm}
\alpha_{\rm H}
&=& \frac{1}{A_4} \left[ \frac{M_{\rm pl}^2}{2}
(F_1-F_2)-(f_1+f_2-2Xf_{2,X}) \right],\label{alHcon}\\
\hspace{-0.5cm}
\alpha_{\rm t}
&=& \alpha_{\rm H}-\frac{2Xf_{2,X}}{A_4}\,.
\label{altcon}
\ea
The difference between $F_1(\phi)$ and $F_2(\phi)$ gives rise to a
non-zero contribution to $\alpha_{\rm H}$.  The term
$f_1+f_2-2Xf_{2,X}$ in Eq.~(\ref{alHcon}) does not generally vanish
for theories beyond the Horndeski domain.  The covariant Galileon
corresponds to $f_1(X)=a_4X^2$ and $f_2(X)=b_4X^2$ with the particular
relation $a_4=3b_4$ \cite{GLPV,Kase14}, in which case
$f_1+f_2-2Xf_{2,X}=(a_4-3b_4)X^2=0$.

In Eq.~(\ref{modelfun}) the constant terms which can exist in the
functions $f_i(X)$ (where $i=1,2$) are assumed to vanish.  In other
words, after the Taylor expansions of the functions $f_i(X)$ around
$X=0$, i.e., $f_i(X)=f_i(0)+f_i'(0)\,X+\mathcal{O}(X^2)$, the
constants $f_i(0)$ have been absorbed to the functions $F_i(\phi)$. As
we will see below, the theories with $F_1(\phi) \neq F_2(\phi)$ are
plagued by the existence of the conical singularity. To avoid this
problem, we need to choose the two constants appearing in $-A_4$ and
$B_4$ are exactly the same as each other.  This corresponds to the
aforementioned tuning of functions required to eliminate the conical
singularity at $r=0$.

\subsection{Theories with $F_1(\phi) \neq F_2(\phi)$}
\label{Fneqsec}

Let us first consider the theories with $F_1(\phi) \neq F_2(\phi)$ and
$f_1(X)=f_2(X)=0$. In this case we have
$\alpha_{\rm H}=\alpha_{\rm t}=F_2(\phi)/F_1(\phi)-1$.  For the
boundary condition $\phi'(0)=0$, the field $\phi$ approaches a
constant value $\phi_c$ as $r \to 0$.  Then the parameters
$\alpha_{\rm H}$ and $\alpha_{\rm t}$ behave as constants around the
origin of a compact object, so the conical singularity cannot be
avoided at $r=0$.

Next, we proceed to the theories with $F_1(\phi) \neq F_2(\phi)$ and
non-vanishing functions of $f_1(X)$ and $f_2(X)$.  An example of
functions $f_1(X)$ and $f_2(X)$, which encompasses both covariant and
extended Galileons, is given by
\be
f_1(X)=a_4X^m, \qquad 
f_2(X)=b_4 X^n\,,
\label{f12}
\ee
where $m, n$ are positive integers ($m,n \geq 1$), and $a_4,b_4$ are constants. 
Since $f_1+f_2-2Xf_{2,X}=a_4X^m+(1-2n)b_4X^n$
and $-2Xf_{2,X}/A_4=4b_4nX^n/[M_{\rm pl}^2F_1(\phi)-2a_4X^m]$ 
in Eqs.~(\ref{alHcon}) and (\ref{altcon}), the parameters 
$\alpha_{\rm H}$ and $\alpha_{\rm t}$ again reduce to 
$\alpha_{\rm H}=\alpha_{\rm t}=F_2(\phi_c)/F_1(\phi_c)-1={\rm constant}$ 
around the origin for the boundary condition $\phi (r \to 0)=\phi_c$.
This constant behavior of $\alpha_{\rm H}$ and $\alpha_{\rm t}$ 
corresponds to the case studied in Sec.~\ref{internalsec}.

We caution, however, that the result in Sec.~\ref{internalsec} has
been derived by assuming that all the terms ${\cal C}_i$ defined in
Eq.~(\ref{Cidef}) vanish.  Let us consider the effect of these terms
around the center of a compact object.  To satisfy the regular
boundary condition $\phi'(0)=0$, we expand the field derivative of the
form
\be
\phi'(r)=\sum_{p=1} \kappa_p\,r^{p}\,,
\label{phiexpand}
\ee
where $\kappa_p$ is a constant and $p$ is a positive integer.  We
substitute Eq.~(\ref{phiexpand}) into Eq.~(\ref{Cidef}) by using the
functions given by Eqs.~(\ref{modelfun}) and (\ref{modelA3}) with
Eq.~(\ref{f12}), under the condition that $\Phi$ approaches a constant
as $r \to 0$. It then follows that
\ba
& &
{\cal C}_1 \to 0\,, \qquad {\cal C}_2 \to 0\,,\nonumber\\
& &
{\cal C}_3 \to 
\kappa_1 M_{\rm pl}^2 e^{-3\Phi}F_{2,\phi}\,,\nonumber\\
& &
{\cal C}_4 \to -2\kappa_1M_{\rm pl}^2e^{-2\Phi}F_{1,\phi}
+8\kappa_1^2a_4e^{-4\Phi}\,,
\label{Ciso}
\ea
where the second term in ${\cal C}_4$ survives only for $m=1$.

Since the terms ${\cal C}_3$ and ${\cal C}_4$ in Eq.~(\ref{Ciso})
approach constants as $r \to 0$, we can absorb these terms into the
cosmological constant $\Lambda$ appearing in $A_2$.  Then, in the
limit that $r \to 0$, Eqs.~(\ref{eq:00}) and (\ref{eq:11}) reduce to
the same forms as those studied in Sec.~\ref{internalsec}, with
non-vanishing constants
$\alpha_{\rm H}=\alpha_{\rm t}=F_2(\phi_c)/F_1(\phi_c)-1$.  Hence the
conical singularity is present at $r=0$ for the theories with
$F_1(\phi) \neq F_2(\phi)$.

We note that the non-analytic function $\phi'(r)=\kappa r^{p}$
($0<p<1$) satisfies the boundary condition $\phi'(0)=0$, but the
second field derivative diverges for $r \to 0$. This gives rise to the
divergent terms $r^{p-1}$ in ${\cal C}_3$ and ${\cal C}_4$, whose
functional dependence is different from that of the term
$2A_4\alpha_{\rm t}/r^2$ in Eq.~(\ref{eq:00}).  This means that such
non-analytic field derivatives do not eliminate the conical
singularity.

\subsection{Theories with $F_1(\phi)=F_2(\phi)$}
\label{Feqsec}

We proceed to the theories satisfying $F_1(\phi)=F_2(\phi)$
with $f_1(X)$ and $f_2(X)$ given by Eq.~(\ref{f12}).
In this case we have 
\ba
\alpha_{\rm H} &=&
\frac{a_4X^m+(1-2n)b_4X^n}{M_{\rm pl}^2F_1(\phi)/2-a_4X^m}\,,\\
\alpha_{\rm t} &=&
\frac{a_4X^m+b_4X^n}{M_{\rm pl}^2F_1(\phi)/2-a_4X^m}\,.
\ea
For the boundary condition $\phi'(0)=0$, both $\alpha_{\rm H}$ and
$\alpha_{\rm t}$ approach 0 in the limit that $r \to 0$.

For the field profile (\ref{phiexpand}) around $r=0$, the quantities
${\cal C}_i$ behave in the same way as Eq.~(\ref{Ciso}) with the
relation $F_1=F_2$. Since ${\cal C}_3$ and ${\cal C}_4$ approach
constants as $r \to 0$, the solutions to Eq.~(\ref{eq:00}) and
(\ref{eq:11}) around the center of a compact body are of the same
forms as Eqs.~(\ref{Phiin1}) and (\ref{Psiin1}), respectively, with
$\alpha_{\rm H}=0$ and $\Lambda$ subject to modifications arising from
constant ${\cal C}_3$ and ${\cal C}_4$.  Since the term
$\alpha_{\rm H}$ vanishes for $r \to 0$, we can avoid the conical
singularity at the origin.

In the above argument the equation of $\phi$ has not been solved
explicitly, but we employed the Taylor expansion of the form
(\ref{phiexpand}) with the boundary condition $\phi'(0)=0$.  In
Sec.~\ref{modelsec} we shall derive solutions to the scalar-field
equation of motion in concrete models under the approximation of weak
gravity and show the existence of the field profile of the form
(\ref{phiexpand}) around the origin.  Provided that the conical
singularity is absent, the solutions derived under the weak-gravity
approximation should not cause any divergent behavior for curvature
quantities.

\section{Models without the conical singularity ($F_1=F_2$)
and Vainshtein mechanism}
\label{modelsec} 

In this section, we solve the equations of motion for the field
$\phi$ and gravitational potentials $\Phi, \Psi$ on the weak
gravitational background characterized by $|\Phi| \ll 1, |\Psi| \ll 1$
in concrete models without the conical singularity ($F_1=F_2$).

We consider the situation in which the dominant contributions to
Eqs.~(\ref{eq:00}) and (\ref{eq:11}) are of the orders of
$A_4\Phi/r^2$ and $A_4\Psi/r^2$ to recover the general relativistic
behavior in the solar system (see Refs.~\cite{KaseHo,DKT} for the
similar approximation).  {}From Eq.~(\ref{eq:33}) the pressure $P_m$
of non-relativistic matter is of the same order as $\rho_m \Psi$, so
$P_m$ is second-order in $\Psi$.  We deal with other terms (including
$\alpha_{\rm H}$ and $\alpha_{\rm t}$) as next-order corrections to
the leading-order terms.

Eliminating the terms $\Phi'$ and $\Psi'$ in Eq.~(\ref{eq:22}) by
using Eqs.~(\ref{eq:00}) and (\ref{eq:11}), we obtain the equation for
$\square \Psi$ coupled to $\square \phi$, where
$\square=d^2/dr^2+(2/r)(d/dr)$.  Differentiating Eq.~(\ref{eq:11})
with respect to $r$ and substituting it into Eq.~(\ref{eq:33}), we can
derive the equation of motion containing the terms $\square \Psi$ and
$\square \phi$.  Eliminating the term $\square \Psi$ by using these
two equations, the resulting scalar-field equation of motion reads
\be
\square \phi \simeq \mu_1 \rho_m+\mu_2\,,
\label{phieq}
\ee
where
\ba
\hspace{-0.2cm}
\mu_1 &=&-\frac{\phi'r(A_{3,X}+\beta)}{4A_4\beta}\,,\label{mu1}\\
\hspace{-0.2cm}
\mu_2 &=&\frac{1}{\beta r}
\Bigg[\left(\frac{A_{2,\phi}}{2}-X A_{2,\phi X}\right)r^2
+(A_{3,\phi}-2X A_{3,\phi X} \nonumber \\
& &+4\phi' XA_{2,XX})r+2\phi'(A_{3,X}+4X A_{3,XX})
-2A_{4,\phi} \nonumber \\
& &+2XA_{4,\phi X}
-\frac{8\phi' XA_{4,XX}+r \alpha_{1}
-4\phi' \alpha_{2}}{r} \Bigg], \label{mu2}
\ea
and 
\ba
\beta
&=& (A_{2,X}+2X A_{2,XX})r+2(A_{3,X}+2X A_{3,XX}) 
\nonumber \\
& &
-\frac{4XA_{4,XX}}{r}
+\frac{2\alpha_{2}}{r}\,,\label{beta} \\
\alpha_1
&=& 2XB_{4,\phi X}-B_{4,\phi}-A_{4,\phi}\,,\\
\alpha_2
&=& 2XB_{4,XX}+B_{4,X}-A_{4,X}\,.
\ea
Here and in the following, the kinetic term $X$ should be understood
as ${\phi'}^2$.  Under the above scheme of approximation,
Eqs.~(\ref{eq:00}) and (\ref{eq:11}) reduce, respectively, to
\ba
\left( r\Phi \right)' &\simeq& -\frac{\rho_m r^2}{4A_4}+\frac{r^2}{4A_4}
\biggl[ A_2-\phi' \left( A_{3,\phi}+
2\phi'' A_{3,X} \right) \nonumber \\
& &+\frac{4\phi'}{r} \left( A_{4,\phi}+
2\phi'' A_{4,X} \right) \biggr]-\frac12 \alpha_{\rm t}\,,\label{gra1}\\
\Psi' &\simeq& \frac{\Phi}{r}-\frac{r}{4A_4} \left( A_2
-2{\phi'}^2 A_{2,X} \right)+\frac{{\phi'}^2  A_{3,X}}{A_4} 
\nonumber \\
& &-\frac{{\phi'}^2  A_{4,X}}{A_4r}+\frac{\alpha_{\rm H}}{2r}\,.
\label{gra2}
\ea

For concreteness, we study the model described by the Lagrangian
(\ref{modelfun}) with the functions
\ba
& &
F_1(\phi)=F_2(\phi) \equiv F(\phi)=e^{-2q \phi/M_{\rm pl}} \,, \nonumber \\
& &
f_1(X)=a_4 X^n\,,\qquad 
f_2(X)=b_4 X^n\,,
\label{model}
\ea
where $q, a_4, b_4$ are constants and $n~(\geq 1)$ is a positive
integer.  Then, the quantities $\mu_1$ and $\mu_2$ appearing in
Eq.~(\ref{phieq}) are given, respectively, by 
\ba
\mu_1 &=&
-\frac{(qM_{\rm pl}F-\beta\phi')r}
{2\beta(M_{\rm pl}^2 F-2a_4{\phi'}^{2n})}\,,\\
\mu_2 &=&
-\frac{4(a_4-b_4)n(2n-1)}{\beta}
\frac{{\phi'}^{2n-1}}{r^2}\,,
\ea
where
\be
\beta=-\frac12 r \left[ 1+4(a_4-b_4)n(2n-1)
\frac{{\phi'}^{2(n-1)}}{r^2} \right]\,.
\label{beta2}
\ee
\subsection{Field profile and gravitational 
potentials around the origin}

We derive the solution to Eq.~(\ref{phieq}) around the center of a
compact body whose density approaches a constant $\rho_m$ as
$r \to 0$.  We search for regular solutions described by
Eq.~(\ref{phiexpand}) around $r=0$ for non-zero values of $q$.

When $n>1$ the terms $\beta \phi'$ and $2a_4{\phi'}^{2n}$ approach 0
in the limit that $r \to 0$, so the quantity $\mu_1$ reduces to
$\mu_1 \to -qr/(2M_{\rm pl}\beta)$.  Integrating Eq.~(\ref{phieq})
with respect to $r$, we obtain the following implicit solution
\be
r^2 \phi'+4n(a_4-b_4){\phi'}^{2n-1}
=\frac{q\rho_m r^3}{3M_{\rm pl}}\,,
\label{implicit}
\ee
where the integration constant is set to 0 to satisfy the boundary
condition $\phi'(0)=0$.  From Eq.~(\ref{implicit}) it is clear that
there is a dominant solution of the form
\be
\phi'(r)=cr\,,
\label{phiso1}
\ee
where $c$ is a constant. 
The coefficient $c$ is different depending on the values 
of $n$, i.e., 
\ba
& & c+8(a_4-b_4)c^3=\frac{q\rho_m}{3M_{\rm pl}}
\qquad ({\rm for}~n=2)\,,\label{cre}\\
& & c=\frac{q\rho_m}{3M_{\rm pl}} 
\qquad ({\rm for}~n>2)\,.
\ea

When $n=1$, the solution to the field equation can be written in terms
of an error function. Expanding it around $r=0$ and using the boundary
condition $\phi'(0)=0$, the dominant solution is given by
\be
\phi'(r)=\frac{q\rho_m}{12M_{\rm pl}(a_4-b_4)}r^3\,.
\label{phiso2}
\ee
Since both Eqs.~(\ref{phiso1}) and (\ref{phiso2}) are of the form
(\ref{phiexpand}), the discussion in Sec.~\ref{Feqsec} to show the
absence of the conical singularity is valid.

Integrating the solution $\phi'(r)=cr^p$ (where $p=1$ for $n\geq 2$
and $p=3$ for $n=1$), it follows that $\phi(r)=\phi_c+cr^{p+1}/(p+1)$
with $\phi_c$ an integration constant.  Then, in the limit that
$r \to 0$, the function $F(\phi)$ converges to a constant value
$F_c \equiv F(\phi_c)$.  In this case the parameters $\alpha_{\rm H}$
and $\alpha_{\rm t}$ are in proportion to $r^{2pn}$, so they vanish at
the origin.  Employing these results and picking up the dominant
contributions to Eqs.~(\ref{gra1}) and (\ref{gra2}) around $r=0$, 
they are integrated to give
\ba
\Phi&=&\frac{\rho_m r^2}{6M_{\rm pl}^2F_c}
-\frac{q c r^{p+1}}{M_{\rm pl}}
+\frac{(1+8q^2F_c)c^2r^{2(p+1)}}{4(2p+3)M_{\rm pl}^2F_c}
\notag\\
&&-\frac{[(4pn+1)a_4+b_4](cr^p)^{2n}}
{(2pn+1)M_{\rm pl}^2F_c}
+\frac{c_{\Phi}}{r}\,,\label{Phiin} \\
\Psi&=&\frac{\rho_m r^2}{12M_{\rm pl}^2F_c}
+\frac{q c r^{p+1}}{(1+p)M_{\rm pl}}
+\frac{(p+2+4q^2F_c)c^2r^{2(p+1)}}{4(p+1)(2p+3)M_{\rm pl}^2F_c}
\notag\\
&&+\frac{[1+(2n-1)p](a_4-b_4)(cr^p)^{2n}}{p(2pn+1)M_{\rm pl}^2F_c}
-\frac{c_{\Phi}}{r}+c_{\Psi}\,, \label{Psiin}
\ea
where $c_{\Phi}$ and $c_{\Psi}$ are integration constants.  We need to
choose $c_{\Phi}=0$ for the regularity of $\Phi$ at $r=0$. 
The constant $c_{\Psi}$, which can be derived by matching solutions at the radius
of a compact body, is finite as in the case of the interior
Schwarzschild solution.  The contributions coming from
$\alpha_{\rm t}$ and $\alpha_{\rm H}$ give rise to terms proportional
to $r^{2pn}$ in Eqs.~(\ref{Phiin}) and (\ref{Psiin}), which vanish as
$r \to 0$.

Evaluating the Ricci scalar $R$ by using the solutions (\ref{Phiin})
and (\ref{Psiin}) with $c_{\Phi}=0$, it follows that $R$ approaches a
finite constant as $r \to 0$. Thus, the model (\ref{model}) is free
from the problem of the conical singularity as a consequence of
vanishing parameters $\alpha_{\rm H}$ and $\alpha_{\rm t}$ at $r=0$.

\subsection{Vainshtein mechanism outside the compact body}

Since there is no conical singularity for the model (\ref{model}) at
the center of a compact object, we proceed to the derivation of
solutions outside the body. In covariant Galileons \cite{cova} with
the dilatonic coupling $F(\phi)=e^{-2q\phi/M_{\rm pl}}$, it is known
that the presence of the terms $f_1(X)=a_4X^2$ and $f_2(X)=b_4X^2$
with $a_4=3b_4$ leads to the recovery of GR inside the so-called
Vainshtein radius $r_V$, even for $|q|$ of the order of 1
\cite{Brax,Kimura,KaseHo}.  Now, we are going to discuss the
Vainshtein mechanism for the model (\ref{model}) {\it without}
imposing the Horndeski condition $a_4+(1-2n)b_4=0$.  For simplicity,
we focus on the theory with $n=2$ and $a_4 \neq 3b_4$.

The Vainshtein radius is characterized by the distance at which the
field derivative in Eq.~(\ref{beta2}) becomes comparable to the first
term in Eq.~(\ref{beta2}), i.e.,
\be
r_V^2 \simeq 24|a_4-b_4|{\phi'}^{2}( r_V )\,, 
\label{rV}
\ee
where $r_V$ can be explicitly known by solving the field equation
(\ref{phieq}).  For the distance $r \gg r_V$ the field
self-interaction term $\mu_2$ is suppressed relative to the term
$\mu_1 \rho_m$, with $\mu_1 \simeq q/M_{\rm pl}$ and
$\beta \simeq -r/2$ \cite{KaseHo}.  Then, Eq.~(\ref{phieq}) is
integrated to give
\be
\phi'(r)=\frac{qM_{\rm pl}r_g}{r^2}\,,\qquad {\rm for}~~r\gg r_V\,,
\label{p1}
\ee
where $r_{g}$ is the Schwarzschild radius of the source defined by
\be
r_{g}\equiv
\frac{1}{M_{{\rm pl}}^{2}}\int_{0}^{r}
\rho_{m} (\tilde{r})\,\tilde{r}^{2}d\tilde{r}\,.
\label{rg}
\ee
Substituting the solution (\ref{p1}) into Eq.~(\ref{rV}), we obtain
the Vainshtein radius
\be
r_V=\frac{(|q|M_{\rm pl}r_g)^{1/3}}{M}\,,\qquad
M \equiv \left( 24|a_4-b_4| \right)^{-1/6}\,,
\label{rV2}
\ee
where $M$ is a constant having a dimension of mass.

For $r_g \ll r \ll r_V$ the field-derivative term $\mu_2$ corresponds
to the dominant contribution to Eq.~(\ref{phieq}), such that
$\mu_1 \rho_m \ll \mu_2 \simeq 2\phi'/r$.  Since the integrated
solution in this regime is $\phi'(r)={\rm constant}$, matching this
solution with Eq.~(\ref{p1}) at $r=r_V$ gives
\be
\phi'(r)=\frac{qM_{\rm pl}r_g}{r_V^2}\,, 
\qquad {\rm for}~~r_g \ll r \ll r_V\,,
\label{p2}
\ee
which is of the same form as that derived in Ref.~\cite{KaseHo} in
Horndeski theories. This reflects the fact that, provided
$a_4 \neq b_4$, the factor $\mu_2$ is independent of the values
$a_4, b_4$ and $n$.  Hence the Vainshtein mechanism can be at work
outside the compact body for the model (\ref{model}) satisfying the
condition $a_4 \neq b_4$.

The integrated solution to Eq.~(\ref{p2}) is given by
$\phi(r)=\phi_0+qM_{\rm pl}r_gr/r_V^2$, where $\phi_0$ is a
constant. Since the second term in $\phi(r)$ is much smaller than
$M_{\rm pl}$ for $|q| \lesssim 1$ and $r_g \ll r \ll r_V$, the
quantity $F(\phi)=e^{-2q\phi/M_{\rm pl}}$ is close to 1 (i.e., the
value of GR) for $|\phi_0| \ll M_{\rm pl}$.  In the following we shall
focus on the case $|\phi_0| \ll M_{\rm pl}$ and employ the
approximation
$A_{4}^{-1} \simeq -(2/M_{\rm pl}^2)(1+2a_4X^2/M_{\rm pl}^2)$.
Integrating Eqs.~(\ref{gra1}) and (\ref{gra2}) after substitution of
the solution (\ref{p2}), we obtain
\ba
\Phi & \simeq & \frac{r_g}{2r}
\biggl[1-2q^2 \left( \frac{r}{r_V} \right)^2
+\frac{q^2(1+8q^2)}{6} \frac{r_g}{r_V} 
 \left( \frac{r}{r_V} \right)^3 \nonumber \\
&&~~~~~-2(a_4+b_4) \frac{M_{\rm pl}^2q^4 r_g^3}
{r_V^8}r \biggr]\,, \label{Phiout}\\
\Psi & \simeq & -\frac{r_g}{2r}
\biggl[1-2q^2 \left( \frac{r}{r_V} \right)^2
-\frac{q^2(1+2q^2)}{3} \frac{r_g}{r_V} 
 \left( \frac{r}{r_V} \right)^3 \nonumber \\
&&~~~~~-8(a_4-b_4) \frac{M_{\rm pl}^2q^4 r_g^3}
{r_V^8} r\ln \frac{r}{r_*} \biggr]\,,\label{Psiout}
\ea
where $r_*$ is an integration constant.  We have dropped the
contribution coming from the term $2a_4X^2/M_{\rm pl}^2$ in
$A_4^{-1}$, as this gives rise to corrections much smaller than the
last terms of Eqs.~(\ref{Phiout}) and (\ref{Psiout}).  For the
distance $r \ll r_V$ the contributions other than the first terms in
the square brackets of Eqs.~(\ref{Phiout}) and (\ref{Psiout}) are much
smaller than 1, under which $\Phi$ and $-\Psi$ are close to the value
$r_g/(2r)$ of GR.

To quantify the deviation from GR, we define the post-Newtonian
parameter
\be
\gamma \equiv -\frac{\Phi}{\Psi}\,.
\ee
The solar-system experiments placed the bound \cite{Will}
\be
\left| \gamma-1 \right|<2.3 \times 10^{-5}\,.
\label{sbound}
\ee
On using the solutions (\ref{Phiout}) and (\ref{Psiout}), 
it follows that 
\ba
\hspace{-0.3cm}
\gamma
&\simeq& 1+\frac{q^2(1+4q^2)}{2}
\frac{r_g}{r_V}\left( \frac{r}{r_V} \right)^3 \nonumber \\
\hspace{-0.3cm}
& &-\frac{2M_{\rm pl}^2 q^4r_g^3r}{r_V^8}
\left[a_4+b_4-4(a_4-b_4) \ln \frac{r}{r_*} \right].
\label{gam}
\ea

The Vainshtein radius $r_V$ depends on the mass scale $M$ defined in
Eq.~(\ref{rV2}).  If the model (\ref{model}) is responsible for the
late-time cosmic acceleration, the parameter $a_4$ is related to the
today's Hubble radius $r_H \sim 10^{28}$ cm as
$|a_4| \sim r_H^4M_{\rm pl}^{-2}$ \cite{DT10}.  Assuming that
$|a_4-b_4|$ is of the order of $|a_4|$, we have
$r_V \sim (|q|r_gr_H^2)^{1/3}$ from Eq.~(\ref{rV2}). Since 
$r_g \sim 10^5$ cm for the Sun, we can estimate 
$r_V \sim 10^{20}$ cm for $|q| \sim 1$. Inside
the solar system ($r \lesssim 10^{14}$ cm), the second term on the rhs
of Eq.~(\ref{gam}) is smaller than $10^{-33}$.  For $r_*$ between
$r_g$ and $r_V$ the term $\ln r/r_*$ is at most of the order of 10, so
the last term on the rhs of Eq.~(\ref{gam}) does not exceed the order
of $10^{-18}$.  Hence the experimental bound (\ref{sbound}) is well
satisfied inside the solar system.

On using the relation $|a_4-3b_4| \sim r_H^4 M_{\rm pl}^{-2}$ and the
solution (\ref{p2}), the parameter $\alpha_{\rm H}$ inside the
Vainshtein radius can be estimated as
\be
|\alpha_{\rm H}| \sim \frac{r_H^4 {\phi'}^4}{M_{\rm pl}^4}
\sim q^4 \frac{r_H^4r_g^4}{r_V^8}\,, 
\ee
which is smaller than $10^{-28}$ for $|q| \lesssim 1$.  Thus, the
deviation from Horndeski theories is significantly suppressed for
$r \ll r_V$.  If the Vainshtein mechanism is {\it not} at work and the
solution is given by Eq.~(\ref{p1}) outside a compact object, we have
that $|\alpha_{\rm H}| \sim q^4r_{\rm H}^4r_g^4/r^8$ and hence
$|\alpha_{\rm H}|$ exceeds the order of 1 for
$r \lesssim \sqrt{|q|r_{\rm H}r_g}$.  For the Sun and $|q| \sim 1$
this condition translates to $r<10^{16}$ cm, so $|\alpha_{\rm H}|$ is
much larger than 1 in the solar system.  The above arguments show how
the Vainshtein mechanism is efficient to suppress both the propagation
of the fifth force and the deviation from Horndeski theories. 

We recall that, for $n=2$, the solution to Eq.~(\ref{phieq}) around
$r=0$ is given by $\phi'(r)=cr$, where $c$ obeys the relation
(\ref{cre}).  For the mass scale $M$ of the order of
$M \sim |a_4|^{-1/6} \sim r_H^{-2/3} M_{\rm pl}^{1/3}$ the term
$8(a_4-b_4)c^3$ is the dominant contribution to the lhs of
Eq.~(\ref{cre}), so the solution around the origin is given by
\be
\phi'(r) \simeq \left( \frac{q\rho_m M^6}{M_{\rm pl}} 
\right)^{1/3}r\,, 
\label{phiinf}
\ee
where we have assumed $q>0$.  In fact, the ratio
$\xi=c/[8|a_4-b_4|c^3]$ is of the order of
$\xi \sim [\rho_0/(q\rho_m)]^{2/3} \ll 1$ for $q \sim 1$, where
$\rho_0 \sim 10^{-29}$ g/cm$^3$ is the cosmological density and
$\rho_m \sim 1$ g/cm$^3$ is the mean density of Sun.

The validity of the solution (\ref{phiinf}) is ensured around $r=0$,
but we can extrapolate it to the surface of the compact body (radius
$r_0$) to estimate the order of $\phi'(r_0)$.  On using the relations
$r_g \sim \rho_m r_0^3/M_{\rm pl}^2$ and
$M=(qM_{\rm pl}r_g)^{1/3}/r_V$ the extrapolation of Eq.~(\ref{phiinf})
gives $\phi'(r_0) \sim qM_{\rm pl}r_g/r_V^2$, which is of the same
order as Eq.~(\ref{p2}).  Thus, the two solutions (\ref{p2}) and
(\ref{phiinf}) smoothly match each other around $r=r_0$.  There are
corrections to the solution (\ref{phiinf}), but they do not change the
order of $\phi'(r)$ inside the body.  This situation is analogous to
what was found in Ref.~\cite{KaseHo} in Horndeski theories.

The smallness of $\alpha_{\rm H}$ in our model comes from the fact that $\alpha_{\rm H}$ depends on $X$ alone ($\alpha_{\rm H} \propto X^2$).
Under the operation of the Vainshtein mechanism, the suppression of the field derivative leads to the small 
value of $\alpha_{\rm H}$.
This is in contrast with the models where $\alpha_{\rm H}$ depends on $\phi$ (i.e., $F_1(\phi) \neq F_2(\phi)$). In the latter case, the models suffer from not only the conical singularity problem, 
but also the breaking of the Vainshtein mechanism 
occurs as shown in Refs.~\cite{Koba,Sakstein,Mizuno}.
The fact that $F_1(\phi)$ and $F_2(\phi)$ are equivalent to 
each other is crucial for the success of the Vainshtein mechanism.

\section{Conclusions}
\label{consec} 

In GLPV theories where the deviation from Horndeski theories is
weighed by the parameter $\alpha_{\rm H}$, we have shown that the
conical singularity arises at the origin of a spherically symmetric
body for nonzero constant $\alpha_{\rm H}$ around the origin.  For
both vacuum and interior Schwarzschild solutions satisfying the
boundary condition $\phi'(r=0)=0$, the Ricci scalar diverges as
$R=-2\alpha_{\rm H}/r^2$ around $r=0$.  In such cases, the spherically
symmetric static body does not exist due to the singularity problem at
its center.  This divergence originates from the nonvanishing
contribution $\alpha_{\rm H}$ to the gravitational potentials, which
does not allow for the recovery of the Minkowski space-time.

For the theories in which $\alpha_{\rm H}$ vanishes as $r \to 0$, it
is possible to avoid the appearance of the conical singularity.  This
requires the condition that the functions $F_1(\phi)$ and $F_2(\phi)$
appearing in Eq.~(\ref{modelfun}) are equivalent to each other. The
functions $f_1(X)$ and $f_2(X)$ need to be chosen such that they do
not involve arbitrary constants which give rise to the difference
between $-A_4$ and $B_4$.  Violation of the condition
$F_1(\phi)=F_2(\phi)$ means that the geometric structure of the
four-dimensional Ricci scalar $R$ is modified, which causes a
geodesically incomplete space-time at $r=0$.

The model described by the functions (\ref{model}), which corresponds
to the extension of covariant Galileons with a dilatonic coupling, is
free from the problem of the conical singularity.  This model is
designed to have the dependence of $\alpha_{\rm H}$ proportional to
$X^n$ around $r=0$. 
We derived the field profile as well as the gravitational potentials
around the center of a compact object under the approximation of weak gravity 
and showed that the Ricci scalar remains finite as a consequence of 
the vanishing $\alpha_{\rm H}$ at the origin. 

For the model (\ref{model}) with $n=2$ we also found that the Vainshtein 
mechanism is at work outside the body to suppress the fifth force and 
the parameter $\alpha_{\rm H}$ in the solar system.  The
regular solution of the field derivative $\phi'(r)$ around the origin
can be extrapolated to match the exterior solution around 
the surface of the body.

There are several issues we have not addressed in this paper. 
First, it will be of interest to study whether the similar properties 
for appearance and disappearance of conical singularities also 
hold for the theories involving the
Lagrangian $L_5$ related to the Einstein tensor.  
Second, we showed that the functional form of Lagrangian must be
properly chosen for eliminating the conical singularity, but 
the absence of a symmetry may give rise to 
quantum corrections which can modify the Lagrangian structure. 
If quantum corrections preserve the structures of {\it four-dimensional} 
curvature quantities like the Ricci scalar $R$, we expect that the modification 
is less harmful. If any difference of a constant between the functions 
$-A_4$ and $B_4$ arises by quantum corrections, this would lead to 
reappearance of the conical singularity.
We leave these interesting issues for future works.

\section*{ACKNOWLEDGEMENTS}
We thank Shinji Mukohyama and Takahiro Tanaka for useful discussions.
ST is supported by the Grant-in-Aid for Scientific Research Fund of
the JSPS No.\,24540286, MEXT Grant-in-Aid for Scientific
Research on Innovative Areas, ``Why does the Universe accelerate? -
Exhaustive study and challenge for the future'' (No. 15H05890),  
and the cooperation program between Tokyo
University of Science and CSIC.



\begin{thebibliography}{10}

\bibitem{review} 
E.~J.~Copeland, M.~Sami and S.~Tsujikawa,
Int.\ J.\ Mod.\ Phys.\ D {\bf 15}, 1753 (2006);
T.~P.~Sotiriou and V.~Faraoni,
Rev.\ Mod.\ Phys.\  {\bf 82}, 451 (2010);
A.~De Felice and S.~Tsujikawa, 
Living Rev.\ Rel.\ {\bf 13}, 3 (2010);
T.~Clifton, P.~G.~Ferreira, A.~Padilla and C.~Skordis, 
Phys.\ Rept.\ \textbf{513}, 1 (2012);
A.~Silvestri and M.~Trodden,
Rept.\ Prog.\ Phys.\  {\bf 72}, 096901 (2009).
 
\bibitem{string}
M.~B.~Green and J.~H.~Schwarz,
Phys.\ Lett.\ B {\bf 149}, 117 (1984). 
 
\bibitem{Horava} 
P.~Horava,
Phys.\ Rev.\ D {\bf 79}, 084008 (2009).

\bibitem{SNIa}
A.~G.~Riess {\it et al.},
Astron.\ J.\  {\bf 116}, 1009 (1998);
S.~Perlmutter {\it et al.},
Astrophys.\ J.\  {\bf 517}, 565 (1999).
 
\bibitem{Planck} 
P.~A.~R.~Ade {\it et al.},
arXiv:1502.01589 [astro-ph.CO]. 

\bibitem{Horndeski} 
G.~W.~Horndeski, 
Int.\ J.\ Theor.\ Phys.\ 10, 363-384 (1974).

\bibitem{Horndeski2} 
C.~Deffayet, X.~Gao, D.~A.~Steer and G.~Zahariade,
Phys.\ Rev.\ D {\bf 84}, 064039 (2011);
T.~Kobayashi, M.~Yamaguchi and J.~'i.~Yokoyama,
Prog.\ Theor.\ Phys.\  {\bf 126}, 511 (2011);
C.~Charmousis, E.~J.~Copeland, A.~Padilla and P.~M.~Saffin,
Phys.\ Rev.\ Lett.\  {\bf 108}, 051101 (2012).

\bibitem{Ost} 
M.~Ostrogradski, 
Mem.\ Ac.\ St.\ Petersbourg VI 4, 385 (1850);
R.~P.~Woodard,
Lect.\ Notes Phys.\  {\bf 720}, 403 (2007).

\bibitem{GLPV} 
J.~Gleyzes, D.~Langlois, F.~Piazza and F.~Vernizzi,
Phys.\ Rev.\ Lett.\  {\bf 114} (2015) 21,  211101.

\bibitem{ADM} 
R.~L.~Arnowitt, S.~Deser and C.~W.~Misner, 
Phys.\ Rev.\ {\bf 117}, 1595 (1960). 

\bibitem{building} 
J.~Gleyzes, D.~Langlois, F.~Piazza, and F.~Vernizzi, 
JCAP \textbf{1308}, 025 (2013).

\bibitem{Hami2} 
J.~Gleyzes, D.~Langlois, F.~Piazza and F.~Vernizzi,
JCAP {\bf 1502}, 018 (2015);
C.~Lin, S.~Mukohyama, R.~Namba and R.~Saitou,
JCAP {\bf 1410}, 071 (2014);
X.~Gao,
Phys.\ Rev.\ D {\bf 90}, 104033 (2014).

\bibitem{Gergely} 
L.~A.~Gergely and S.~Tsujikawa,
Phys.\ Rev.\ D {\bf 89}, 064059 (2014).

\bibitem{Kase14} 
R.~Kase and S.~Tsujikawa,
Phys.\ Rev.\ D {\bf 90}, 044073 (2014).

\bibitem{Koyama} 
A.~De Felice, K.~Koyama and S.~Tsujikawa,
JCAP {\bf 1505}, 058 (2015).

\bibitem{Tsuji15} 
S.~Tsujikawa,
Phys.\ Rev.\ D {\bf 92}, 044029 (2015).
  
\bibitem{Will} 
C.~M.~Will,
Living Rev.\ Rel.\  {\bf 9}, 3 (2006).

\bibitem{Kimura} 
R.~Kimura, T.~Kobayashi and K.~Yamamoto,
Phys.\ Rev.\ D {\bf 85}, 024023 (2012);
K.~Koyama, G.~Niz and G.~Tasinato,
Phys.\ Rev.\ D {\bf 88}, 021502 (2013).

\bibitem{KaseHo} 
R.~Kase and S.~Tsujikawa,
JCAP {\bf 1308}, 054 (2013).

\bibitem{Vainshtein} 
A.~I.~Vainshtein,
Phys.\ Lett.\ B {\bf 39}, 393 (1972).

\bibitem{Koba} 
T.~Kobayashi, Y.~Watanabe and D.~Yamauchi,
Phys.\ Rev.\ D {\bf 91}, 064013 (2015).

\bibitem{Sakstein}
K.~Koyama and J.~Sakstein,
Phys.\ Rev.\ D {\bf 91}, 124066 (2015).
    
\bibitem{Mizuno}
R.~Saito, D.~Yamauchi, S.~Mizuno, J.~Gleyzes 
and D.~Langlois,
JCAP {\bf 1506}, 008 (2015).

\bibitem{KTD15} 
R.~Kase, S.~Tsujikawa and A.~De Felice,
arXiv:1510.06853 [gr-qc] [Phys.\ Rev.\ D (to be published)].

\bibitem{Gergely2} 
R.~Kase, L.~A.~Gergely and S.~Tsujikawa,
Phys.\ Rev.\ D {\bf 90}, 124019 (2014).

\bibitem{Bellini} 
E.~Bellini and I.~Sawicki,
JCAP {\bf 1407}, 050 (2014).

\bibitem{Moto} 
T.~Kobayashi, H.~Motohashi and T.~Suyama,
Phys.\ Rev.\ D {\bf 85}, 084025 (2012).

\bibitem{Khoury} 
J.~Khoury and A.~Weltman,
Phys.\ Rev.\ D {\bf 69}, 044026 (2004).

\bibitem{DKT} 
A.~De Felice, R.~Kase and S.~Tsujikawa,
Phys.\ Rev.\ D {\bf 85}, 044059 (2012).

\bibitem{Brans} 
C.~Brans and R.~H.~Dicke,
Phys.\ Rev.\  {\bf 124}, 925 (1961).

\bibitem{Gas} 
M.~Gasperini and G.~Veneziano,
Astropart.\ Phys.\  {\bf 1}, 317 (1993).

\bibitem{cova} 
C.~Deffayet, G.~Esposito-Farese and A.~Vikman,
Phys.\ Rev.\ D {\bf 79}, 084003 (2009).

\bibitem{exga}
A.~De Felice and S.~Tsujikawa,
JCAP {\bf 1202}, 007 (2012);
A.~De Felice and S.~Tsujikawa,
JCAP {\bf 1203}, 025 (2012).

\bibitem{Brax} 
P.~Brax, C.~Burrage and A.~C.~Davis,
JCAP {\bf 1109}, 020 (2011).

\bibitem{DT10}
A.~De Felice and S.~Tsujikawa,
Phys.\ Rev.\ Lett.\  {\bf 105}, 111301 (2010).

\end{thebibliography}
\end{document}